\documentclass[useAMS,usenatbib]{mn2e}

\usepackage {graphicx}
\usepackage {times}

\title[A new QSO at z=5.96]{Discovery of a new high redshift QSO at z=5.96 with the Subaru telescope\footnotemark[0]\thanks{Based  on data collected at Subaru Telescope, which is operated by the National Astronomical Observatory of Japan.}}
\author[T. Goto]{Tomotsugu Goto
 \thanks{E-mail:tomo@ir.isas.jaxa.jp} 
\\
 Institute of Space and Astronautical Science,  Japan Aerospace Exploration Agency,
 3-1-1 Yoshinodai, Sagamihara, Kanagawa 229-8510, Japan}
\begin{document}

\date{\today; in original form 2006 March 22}

\pagerange{\pageref{firstpage}--\pageref{lastpage}} \pubyear{2006}

\maketitle

\label{firstpage}

\begin{abstract}
We report a discovery of a new high redshift quasar at $z=5.96$, observed with the FOCAS long-slit spectrograph on board the Subaru telescope. 
The spectrum shows strong and broad Ly$\alpha$+NV emission lines with a sharp discontinuity to the blue side. A Ly$\beta$+OVI emission line is also detected, providing a consistent redshift measurement with the Ly$\alpha$+NV emission.  The QSO has an absolute magnitude of $M_{AB,1450}=-26.9$ ($H_{0}=50$km s$^{-1}$ Mpc$^{-1}$, $q_0=0.5$).
 The spectrum shows significant flux in the region 8000-8300 $\AA$ and thus does not show a complete  Gunn-Peterson trough in the redshift range 5.58 to 5.82, along the line of sight to this $z=5.96$
 QSO. Therefore the Universe was already highly ionized at $z=5.82$. 
\end{abstract}


\section{Introduction}

\begin{figure*}
\begin{center}
\includegraphics[scale=1]{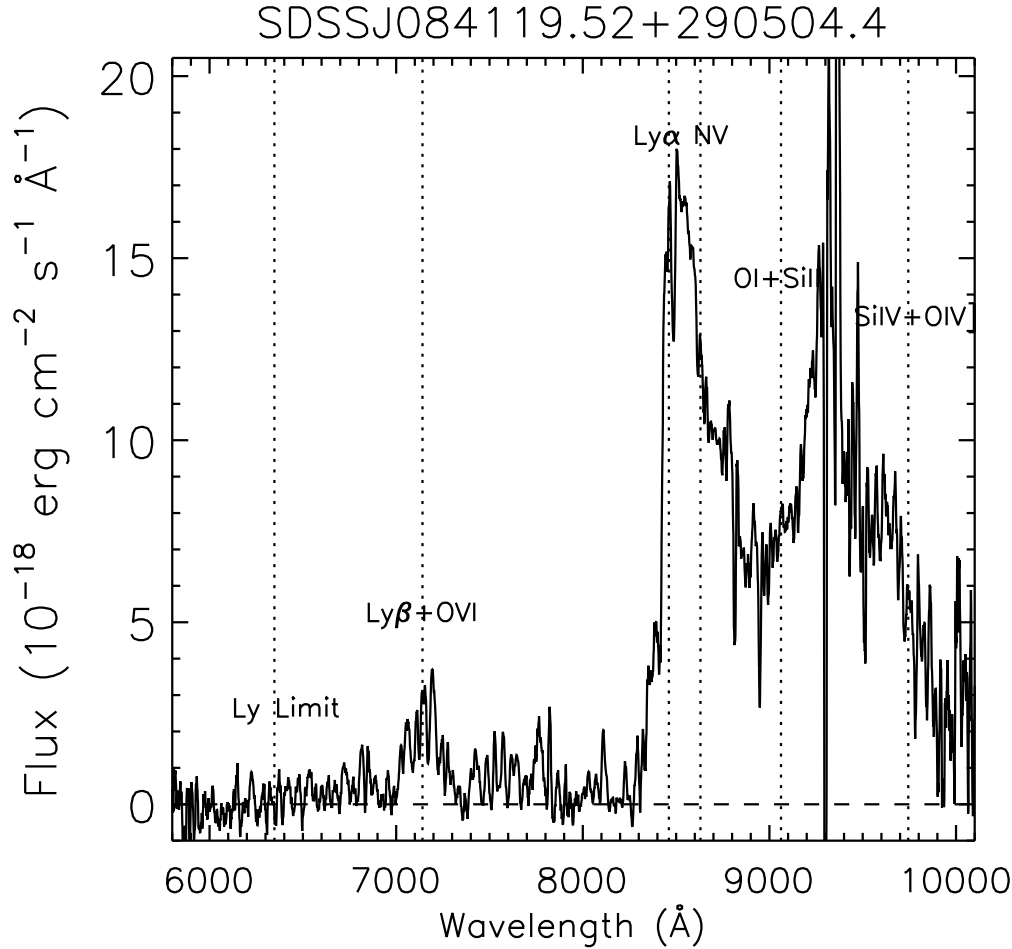}
\end{center}
\caption{
The Subaru/FOCAS spectrum of a QSO, SDSSJ084119.52+290504.4, at z=5.96. The spectrum is smoothed using a 10 \AA\ box. Expected locations of various emission lines at $z=5.96$ are indicated with the dotted line. 
}\label{fig:QSO_spectrum}
\end{figure*}

\begin{figure*}
\begin{center}
\includegraphics[scale=0.85]{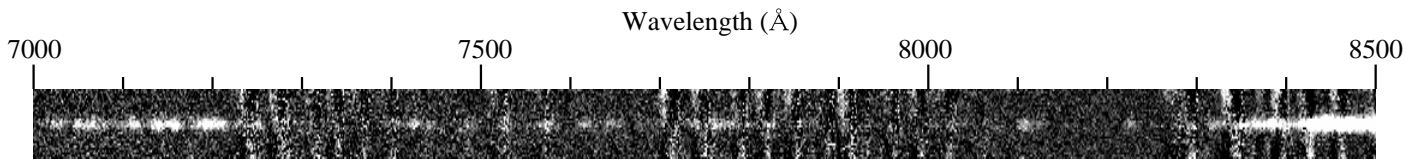}
\end{center}
\caption{
The sky-subtracted two-dimensional spectrum of the QSO. The width of the vertical axis is 30 pixel or 9 arcsec. The vertical axis is expanded by 50\% for clarity. 
}\label{fig:QSO_2d}
\end{figure*}

 
High-redshift QSOs are highly important objects in various scientific aspects.
They provide direct probes of the epoch
when the first generation of galaxies and QSOs formed.
The absorption spectra of these QSOs reveal the state of the intergalactic 
medium (IGM) close to the reionization epoch \citep{1999ApJ...519..479H,2000ApJ...542L..69M,2000ApJ...542L..75C}.
The lack of a Gunn-Peterson trough 
\citep{1965ApJ...142.1633G} 
 in the spectrum of the luminous QSO at $z=6.43$ \citep{Fan_z6.43,2003AJ....126....1W,2005AJ....129.2102W} indicates that the universe was already highly ionized
at that redshift.
Assuming that the QSO is radiating at the Eddington luminosity, 
this object contains a central black hole of several billion solar masses.
The assembly of such massive objects in a timescale shorter than
1 gigayear yields constraints on models of the formation of massive black holes 
\citep[e.g.,][]{2001ApJ...552..459H}.
The abundance and evolution of such QSOs can provide sensitive
tests for models of QSO and galaxy evolution. 

 The Sloan Digital Sky Survey \citep[SDSS,][]{2000AJ....120.1579Y} uses
a dedicated 2.5m telescope and a large format CCD camera 
to obtain images in five broad bands 
\citep[$u$, $g$, $r$, $i$ and $z$ centred at 3551, 4686, 6166, 7480 and 8932 \AA, respectively;][]{1996AJ....111.1748F} 
over 10,000 deg$^2$ of high Galactic latitude sky.
 This unprecedented large sky coverage 
 provides us a unique
 opportunity to find a very rare class of objects such as highest
 redshift QSOs, passive spiral galaxies\citep{2003PASJ...55..757G}, and E+A galaxies \citep{2003PASJ...55..771G,2004A&A...427..125G,2005MNRAS.357..937G}. The inclusion of the reddest band, $z$, in principle enables the discovery of QSOs up to $ z \sim 6.7$ from the SDSS data as a $z$-band only detection \citep{Fan2000}. In this work, we have used the fourth public data release of the SDSS \citep{2006ApJS..162...38A}.

Here we report our pilot search for high redshift QSOs with the Subaru telescope and a successful discovery of a new QSO at $z=5.96$.

\section{Target Selection}\label{data}

 Our target selection is similar to that is used by  \citet{Fan2000,Fan2001iz22,2001AJ....121...31F,Fan_z6.43}.
 At $z>5.7$, Ly$\alpha$ emission of QSOs moves out of the $i$-band and into $z$-band, the reddest filter of the SDSS, and no flux shall be found in bluer $u,g,r,$ and $i$-bands. Therefore, we specifically target faint point sources detected only in $z$-band in the SDSS imaging data ($i_{AB}>22.0$ and $z_{AB}<20.7$)\footnote{We express $u,g,r,i$ and $z$ magnitudes in AB system, and $J$ magnitude in Vega system throughout this paper.}. In addition, we require targets to have $i_{AB}-z_{AB}>2.2$ to avoid contamination from late-type stars \citep{1999ApJ...522L..61S,2000ApJ...531L..61T}.  Previously, such a target selection successfully found the currently highest redshift QSO at $z=6.43$ \citep{Fan_z6.43}, and is known to work quite well \citep{2000AJ....120.1607Z,Fan2001iz22,2005AJ....130...13C}. 
 
  For these candidates, we have obtained $J$-band imaging using one night of the UKIRT and three half nights of the APO3.5m telescope time to further remove contaminations from cosmic-rays and late-type stars. 
We have selected our targets from the fourth public data release of the SDSS \citep[][6670 deg$^2$ in total]{2006ApJS..162...38A}. 
In total, 131 candidates were observed in $J$-band. The $z-J$ colour is a powerful separator of high-z QSOs (blue in $z-J$) from late-type stars \citep[red in $z-J$;][]{Fan2001iz22} since the continuum shape of QSOs is relatively flat toward near infrared, while the flux of late-type stars continues to rise toward longer wavelengths. We only target those candidates whose $z-J$ colour is consistent to be high-$z$ QSOs ($z_{AB}-J_{Vega}\leq1.7$) for spectroscopic follow-up with the Subaru telescope.

\section{Observation}
\begin{table*}
 \centering
 \begin{minipage}{140mm}
  \caption{Quasar properties}\label{tab:QSO}
  \begin{tabular}{@{}clrrrrlrlr@{}}
  \hline
   Object & Redshift & $i_{AB}^*$ & $z_{AB}^*$ & $J_{Vega}$ \footnote{Observed with the UKIRT}  & $M_{AB,1450}$ &  & \\
 \hline
SDSSJ084119.52+290504.4 & 5.96 & 22.54 & 19.84 & 18.9 & $-$26.86 &  &  &  & \\
\hline
\end{tabular}
\end{minipage}
\end{table*}

 We have spectroscopically observed 26 candidates that satisfied the criteria in Section \ref{data} with the  FOCAS \citep[The Faint Object Camera and Spectrograph;][]{2002PASJ...54..819K} mounted on the Subaru telescope on the night of February 3rd, 2006. The weather was fine all through the night except for relatively strong wind that blurred the seeing to be  $\sim$1.5 arcsec.
 We used 300 lines mm$^{-1}$ grating and the 058 order cut filter with 0.8'' wide slit, giving a spectroscopic resolution of $R\sim600$. This setting will give us wavelength coverage of 6000\AA\ to 10000\AA\ .    
    The spatial resolution used was 0.3'' pixel$^{-1}$ by 3-pixel, on-chip binning. 
 Data reduction was carried out using the standard IRAF routines. We have used the Ar lamp to calibrate wavelength. We observed standard stars, G191B2B and HZ44, with the 2'' slit-width for a flux calibration. 
 Under the condition of 1.5'' seeing, we estimate that the slit-loss of the 0.8'' slit is 47\% compared to the 2'' slit assuming the Gaussian PSF, and corrected the flux calibration by this factor. However, we caution readers not to strongly interpret the flux calibration due to the uncertainties from the telescope pointing, tracking, time-variability of the seeing size and so on.

 Out of the 26 candidates, we have found one QSO. The rest of the targets were either M or L dwarf stars.
 We discuss the details of this QSO in the next section.

\section{A new QSO at z=5.96}\label{QSO}

 We present the basic information of the QSO in table \ref{tab:QSO}. 
 We show the observed spectrum in Figure \ref{fig:QSO_spectrum}. 
 Three exposures of 1200 sec each were combined. 
 The spectrum is smoothed using a 10 \AA\ box. 
 The spectrum shows an unambiguous signature of a high redshift QSO. The broad and strong Ly$\alpha$ +NV emission lines are present at 8465\AA\ with a sharp discontinuity to the blue side due to the strong absorption by neutral hydrogen.  The  Ly$\alpha$  and NV emission lines are blended. A straight Gaussian fit to the line suggests that this QSO is at $z=5.96$. The Ly$\alpha$ shows self-absorption at 8487\AA.  Ly$\beta$+OVI emission lines are detected  at 7141 \AA\, providing a consistent redshift measurement of $z=5.96$.  This is the 11th highest redshift QSO known to date. 
 The spectrum shows no detectable flux at $<6350\AA$ because of the Lyman limit system. 
 At the redward of the Ly$\alpha$, there is a slight sign of OI+SII($\lambda$1302) and SiIV+OIV]($\lambda$1400) emissions. They are, however, not clearly detected partly due to the increased noise on the spectrum, and possibly because of the metals are not produced yet at this high redshift.  The equivalent width measurement of Ly$\alpha$+NV line is difficult due to the uncertain continuum level. Using the continuum level determined from the redward of the line, we measured restframe equivalent width of $\sim$58\AA, quite typical of lower redshift QSOs. The measurement of Ly$\beta$+OVI line is difficult as well. We measure restframe equivalent width of $\sim$46\AA. However, these numbers are highly uncertain. 
 The extinction corrected absolute magnitude at 1450\AA\ is $M_{AB,1450}=-26.86$ (We used $H_{0}=50$ and $q_0=0.5$ for comparison purpose), assuming that  the QSO is not gravitationally magnified.  This is a typical luminosity of $z\sim 6$ QSOs \citep[c.f.,][]{2001AJ....121...31F}. It is important to test the gravitational amplification with a high-resolution imaging \citep[see ][]{2002PASJ...54..975S}. 
 The QSO is not detected in the FIRST radio survey \citep{1995ApJ...450..559B} at the 1mJy level in 20cm. 
 The features around 9300\AA\ are affected by the residuals in background sky subtraction and are not real. There are possible signs of absorption lines at 8810,8946,9509 \AA. However, the locations of these lines coincides with strong atmospheric features, and thus, the detection is questionable. 

 We show the sky-subtracted two-dimensional spectrum in Figure \ref{fig:QSO_2d}. In the spectrum, there remains detectable flux blueward of the Ly$\alpha$ emission ($7300\AA<\lambda<8300\AA$). (For example, the fluxes around 8226,8109 and 7612\AA are at the relatively clear regions of the spectrum).
This suggests that the Gunn-Peterson trough is not complete and the universe is already highly ionized at $z=5.96$.



\section{Conclusions}

We have found a new QSO at $z=5.96$ from the SDSS data. The object was selected for a spectroscopic follow-up with the Subaru telescope as a $z$-band only detection with blue $z-J$ colour. The broad Ly$\alpha$+NV emission and the Ly$\beta$+OVI emission provides a consistent redshift measurement of $z=5.96$, making this the 11th highest redshift QSO to date. 
 The spectrum shows significant flux in the region 8000-8300 $\AA$ and thus does not show a complete  Gunn-Peterson trough in the redshift range 5.58 to 5.82, along the line of sight to this $z=5.96$
 QSO. Therefore the Universe was already highly ionized at $z=5.82$. 
 It is our important future work to compute surveyed area and space density of high redshift QSOs once we finish observing all the selected targets.

\section*{Acknowledgments}

We thank Dr. M.Yagi, Y.Ohyama and C.Yamauchi for some helpful suggestions, and Dr. T.Hattori for friendly help during the observation. We thank the anonymous referee for many insightful comments, which significantly improved the paper.

 The United Kingdom Infrared Telescope is operated by the Joint Astronomy Centre on behalf of the U.K. Particle Physics and Astronomy Research Council.

 Use of the UKIRT 3.8-m telescope for the observations is supported by NAOJ.

 This research is in part based on observations obtained with the Apache Point Observatory 3.5-meter telescope, which is owned and operated by the Astrophysical Research Consortium.
   
    Funding for the creation and distribution of the SDSS Archive has been provided by the Alfred P. Sloan Foundation, the Participating Institutions, the National Aeronautics and Space Administration, the National Science Foundation, the U.S. Department of Energy, the Japanese Monbukagakusho, and the Max Planck Society. The SDSS Web site is http://www.sdss.org/.

    The SDSS is managed by the Astrophysical Research Consortium (ARC) for the Participating Institutions. The Participating Institutions are The University of Chicago, Fermilab, the Institute for Advanced Study, the Japan Participation Group, The Johns Hopkins University, Los Alamos National Laboratory, the Max-Planck-Institute for Astronomy (MPIA), the Max-Planck-Institute for Astrophysics (MPA), New Mexico State University, University of Pittsburgh, Princeton University, the United States Naval Observatory, and the University of Washington.



\label{lastpage}

\end{document}